\documentstyle[aps,prb,preprint]{revtex}
\begin{document}

\draft

\title{Realistic Calculations of Correlated Incompressible Electronic States in
GaAs--$\mbox{Al}_{x} \mbox{Ga}_{1-x}$As Heterostructures and Quantum Wells}

\author{M. W. Ortalano}
\address{Department of Physics, University of Maryland, College Park, Maryland
20742}

\author{Song He}
\address{AT\&T Bell Laboratories, Murray Hill, NJ 07974}

\author{S. Das Sarma}
\address{Department of Physics, University of Maryland, College Park, Maryland
20742}

\date{\today}

\maketitle

\begin{abstract}
    We perform an exact spherical geometry finite-size diagonalization
calculation for the fractional quantum
Hall ground state in three different experimentally relevant
GaAs-$\mbox{Al}_{x} \mbox{Ga}_{1-x}$As systems: a
wide parabolic quantum well, a narrow square quantum well, and a
heterostructure. For each system we obtain the Coulomb pseudopotential
parameters
entering the exact diagonalization calculation by using the realistic subband
wave function from a self-consistent electronic structure
calculation within the local density approximation (LDA) for a range of
electron
densities.
We compare our realistic LDA pseudopotential parameters with those from widely
used simpler model approximations in order to estimate the accuracies of the
latter.
We also calculate the overlap between the exact numerical ground state and the
analytical Laughlin state as well as the
excitation gap as a function of density. For the three physical systems we
consider the calculated
overlap is found to be large in the experimental electron density range. We
compare
our calculated excitation gap energy to the experimentally obtained activated
transport energy gaps after subtracting out
the effect of level broadening due to collisions. The agreement between our
calculated excitation gaps and the experimental measurements is excellent.

\end{abstract}

\vfill

\newpage

\section{Background}

\indent  The fractional quantum Hall effect (FQHE) has been observed in
high mobility GaAs-$\mbox{Al}_{x} \mbox{Ga}_{1-x}$As quantum structures at
low temperatures and in strong magnetic fields
\cite{TSUI,STORM,CHANG,SHAYEGAN,WILLET1,BOEB,WILLET2}. This effect produces
quantized
plateaus in the Hall resistivity concurrent with minima in the longitudinal
resistivity at special values of the electron density. A well formed electron
correlation driven energy
gap separating the ground state from the excited states occurring at these
special
 densities and magnetic fields is the underlying reason for the FQHE
phenomenon. The special density and magnetic field needed for the FQHE
correspond to a Landau level filling factor $ \nu=p/q$ where q is an odd
integer in the simplest
situation.
Laughlin's theory \cite{LAUGHLIN} of the FQHE, valid for filling fractions of
the form
$\nu = 1/m$, where $m$ is an odd integer, is based on the following
two-dimensional (2D) many-body
wave function:

\begin{equation}
\psi _{m} \left(z_{1}, \ldots ,z_{N} \right) = \prod_{i<j} \left(z_{i}-z_{j}
\right)^{m} \exp \left(-1/4 \sum_{j=1}^{N} |z_{j}|^{2} \right)
\end{equation}

\noindent where $z_{i}=x_{i}-iy_{i}$ is the complex representation of the
$i^{th}$ electron's 2D position vector.
The Laughlin state describes a droplet of an incompressible correlated 2D
electron liquid. At
these special values of the filling fraction, the system has an excitation gap
that separates the ground state from the excited states. The elementary
excitations are fractionally charged anyons.
In a single layer 2D system, Laughlin's theory explains the FQHE at
$\nu=1/m$ and $\nu=1-1/m$, where m is an odd integer. The hierarchy
construction
extends this theory to filling fractions of the form $\nu=p/q$ where $q$ is an
odd integer \cite{PRANGE,HALPERIN1,HALDANE3,HALPERIN2,JAIN}.

\indent Direct numerical calculations involving the exact diagonalization of a
small ($4 \sim 10$) number of interacting 2D electrons has verified the
Laughlin theory extremely well.
The geometry of choice in most numerical simulations is the
translationally invariant, rotationally invariant spherical geometry
\cite{HALDANE3}. Using
just a small number of electrons, $N_{e}$, ($N_{e} \le 10$) can yield an
accurate picture of the physical system. In this geometry, a magnetic
monopole is placed at the center of a sphere of radius $R$, producing a
radial magnetic field $B$. By the Dirac quantization condition, the total
flux $4 \pi R^{2} B$ must be an integral multiple, $2S$, of the elemental
flux $hc/e$. From the hierarchy construction a relationship between $2S$
and the filling factor $\nu$ can be found. For $\nu=1/3$, $2S=3(N_{e}-1)$.
The appropriate unit of length is $l_{c}=\sqrt{ \hbar c/e B}$.
If all of the electrons in the system are placed in the lowest Landau
level and the coupling between Landau levels is ignored the resulting many-body
problem is exactly soluble. The Hilbert space is of finite dimensions and an
exact finite dimensional Hamiltonian matrix can be written down. This matrix
can be diagonalized by standard techniques to find the energy eigenvalues
and the eigenvectors. It is for this reason that exact finite size
diagonalization
 has become such an important tool for studying the FQHE.
In fact such numerical studies have been instrumental
in confirming Laughlin's many-body wave function \cite{FANO,HALDANE1,HALDANE4}.

\indent  One direction that research in finite size diagonalization studies of
the FQHE has taken is in increasing
system size, {\em i.e.} increasing the number of electrons in the numerical
simulation. In this
paper, we take a different approach and place the emphasis on quantitatively
improving the model used in describing the
 electron-electron interaction in the system to make it more realistic with
respect to the experimental systems.
Historically the first studies
of the FQHE used a pure 2D Coulomb potential, and the pure $1/r$ Coulomb
interaction, where $r$ is the separation between 2D electrons,
is still the most popular model for exact finite size diagonalization studies.
It was later pointed out that the finite
layer thickness in quantum structures would cause the short range part of
the Coulomb interaction to become softened. This could cause the ground state
of the system to be no longer incompressible for very thick layers, and
the FQHE may eventually be destroyed in thick layers. Zhang and Das Sarma
\cite{SDS5} and
He {\em et al.} \cite{SONG} investigated this `finite thickness'
phenomena using a simple variant of the Coulomb interaction, namely

\begin{equation}
V(\vec{r})=\frac{e^{2}}{\kappa} \frac{1}{ \sqrt{r^{2}+ \lambda ^{2}}}
\end{equation}

\noindent where the length scale $\lambda$ represents the finite extent of the
electron wave
function in the $z$ direction and $\vec{r}$ is the 2D position vector. Recently
the FQHE has been studied using more sophisticated models for the
electron-electron
interaction \cite{SDS5,SONG,BELKHIR2}. The most accurate approximation that one
can make in this respect is
to do a self-consistent electronic structure calculation within the framework
of the local density approximation (LDA) to describe the interaction. In this
paper
we perform such a calculation for a wide parabolic quantum well (PQW), a narrow
square quantum well (SQW), and a heterostructure, and we use the realistic LDA
electronic structure to compute the effective electron-electron interaction
entering the exact diagonalization study.
We use Haldane's spherical geometry \cite{HALDANE3} to do a fully spin
polarized finite size FQHE diagonalization calculation for six electrons, which
should be sufficient for our purpose.
The many-body Hamiltonian matrix for the FQHE is constructed using these
LDA-Coulomb matrix elements.
Using the Lanczos method, we calculate the eigenvalues
and the eigenvectors of this fairly sparse FQHE Hamiltonian matrix, following
standard techniques.
The overlap with the Laughlin $\nu =1/3$ state is found by diagonalizing
the Hamiltonian matrix using the pseudopotential parameters appropriate for
this
 state and then calculating the inner product between this state and the
exact numerical ground state found
using the LDA pseudopotential parameters.

\indent The excitation gap is determined by looking at the size of the cusp
at the relevant filling factor. In order to compare this calculated excitation
gap to relevant
experimental transport data, we subtract out the level broadening due to
collisions,
$\Gamma$. Using $\mu$, the experimentally determined value for the mobility,
$\Gamma$ can be determined from

\begin{equation}
\Gamma = \left( \frac{\hbar}{2 \tau_{s}} \right) = \left( \frac{\hbar e}{2m^{*}
\mu} \right).
\end{equation}

\noindent  However, this equation is not quite correct \cite{SDS2} for the
single particle level broadening if the scattering is strongly
peaked in the forward direction as it is in modulation doped GaAs structures.
We estimate the correct $\Gamma$ ({\em i.e.} the single particle broadening)
using the transport data by employing a simple theory \cite{SDS2} which
accounts
for this forward scattering correction.
We then take the calculated excitation gap and subtract out $2 \Gamma$
when comparing with experiment.

\section{Local Density Approximation}

\indent We begin with a description of the procedure for a self-consistent
electronic structure calculation at zero temperature and in zero magnetic
field. We assume that the effective mass and the static dielectric
constant do not vary over the width of the quantum well and that interface
grading and dielectric mismatch are negligible.

\indent We solve the effective-mass, single-particle effective Schr\"{o}dinger
equation for a particle in a quantum well and Poisson's equation for the
electrostatic potential due to the free electric charge self-consistently.
Assuming separability of the planar and perpendicular degrees of freedom,
the three dimensional Schr\"{o}dinger equation reduces to

\begin{equation}
\left( -\frac{\hbar ^{2}}{2m^{*}} \frac{d^{2}}{dz^{2}} + V_{EFF}(z) \right) \xi
_{i}(z) = E_{i} \xi _{i}(z)
\end{equation}

\noindent where $\xi_{i}(z)$ and $E_{i}$ are the subband wave functions and
energies respectively, and the effective one electron potential energy
$V_{EFF}(z)$ is given by

\begin{equation}
\label{potential}
V_{EFF}(z) = V_{W}(z) + V_{H}(z) + V_{XC}(z)
\end{equation}

\noindent with $V_{W}(z)$ being the quantum well confinement potential,
$V_{H}(z)$ the self-consistent Hartree potential, and $V_{XC}(z)$ is the
exchange-correlation potential.
For the bare confining potential of a quantum well of width $a$, we take

\begin{eqnarray}
V_{W}(z) & = & \left \{ \begin{array}{ll}
                   V_{0} \theta \left(|z| -a/2 \right),  & \mbox{ for a SQW}\\
                   V_{0} \theta \left(|z|-a/2 \right)+ \alpha z^{2} \theta
\left( a/2 - |z| \right ) ,    & \mbox{ for a PQW}
                  \end{array}
                 \right.
\end{eqnarray}

\noindent with $V_{0}$ being the barrier height for a square quantum well and
the barrier height above the edge of the parabolic portion for a
parabolic quantum well, and $\alpha$ is the
curvature of the parabolic quantum well.

\indent Poisson's equation for the Hartree potential is given by

\begin{equation}
\frac{d^{2}V_{H}(z)}{dz^{2}}=-\frac{4\pi e^{2}}{\kappa} \left( n(z)-n_{I}(z)
\right)
\end{equation}

\noindent where $\kappa$ is the background dielectric constant for GaAs, $n(z)$
is the electron density computed from the effective single particle subband
wave functions and $n_{I}(z)$ is the density of donor impurities. We do not
include
$n_{I}(z)$ explicitly but include it via the boundary conditions in the
solution of Poisson's equation.

\indent The areal density is determined from the subband wave functions by

\begin{equation}
n(z) = 2 \sum_{i}^{imax} N_{i} | \xi _{i}(z)|^{2}
\end{equation}

\noindent where $N_{i}$ is the occupancy of the $i$th subband and is given by

\begin{equation}
N_{i}= \int \frac{kdk}{2 \pi} \theta \left( E_{F}-E_{i} - \frac{k^{2}}{2m^{*}}
\right)= \frac{m^{*}}{2 \pi} \left( E_{F} -E_{i} \right) .
\end{equation}

\noindent The chemical potential $E_{F}$ is determined by the relation

\begin{equation}
N_{s} = \int dz n(z) = 2 \sum_{i}^{imax} N_{i} = \frac{1}{2 \pi}
\sum_{i}^{imax} 2m^{*} \left( E_{F}-E_{i} \right)
\end{equation}

\noindent where $N_{s}$ is the total surface density. The above equation is
inverted
to give $E_{F}$ and $imax$.

\indent Many-body effects beyond the Hartree approximation are included
by means
of the density functional theory in the local density approximation (LDA)
\cite{HK,KS,SK}.
A chief concern of density functional theory is the calculation of the
exchange-correlation energy functional, $V_{XC}[n]$. This functional of the
electron
density contains all those interaction parts of the energy functional which in
general
are unknown. The local density approximation consists of replacing the
functional $V_{XC}[n]$ with a function $V_{XC}(n)$ whose value at a given
point in space $z_{0}$, where the density is $n(z_{0})$, is determined
as though the density was constant and equal to $n(z_{0})$ everywhere.
The validity of this approximation requires that the variation of the
electron density be small over distances of the order of a Fermi wavelength.
This condition is
in general violated in most semiconductor quasi-2D systems. However, there
is considerable evidence that this approximation when used in these
systems gives excellent agreement with experiment
\cite{SDS1,ANDO1,ANDO2,ANDO3,SDS4}.
For the exchange-correlation potential, we
used the parametrization due to Hedin and Lundqvist \cite{HL}:

\begin{equation}
V_{XC}(z) = - \left( 1+0.7734 x \ln (1+x^{-1}) \right) \left( \frac{2}{\pi
\beta r_{s}} \right) Ry^{*}
\end{equation}

\noindent where $\beta=(4/9 \pi)^{1/3}$, $x=r_{s}/21$ and

\begin{equation}
 r_{s} = \left( \frac{4}{3} \pi a^{*^{\scriptstyle 3}} n(z) \right) ^{-1/3}
\end{equation}

\noindent with $a^{*}$ and $Ry^{*}$  being the effective Bohr radius and the
effective Rydberg respectively in GaAs.

\indent The self-consistent procedure is to start with an initial guess for the
electron density $n(z)$. The Hartree and exchange-correlation potentials are
then computed for this density. The Schr\"{o}dinger equation is then solved
numerically to obtain $\xi_{i}$ and $E_{i}$. A new density is then computed
and compared to the previous $n(z)$ through

\begin{equation}
\eta = \frac{\int dz \left| n_{new}(z) - n_{old}(z) \right|}{\int dz n_{old}}.
\end{equation}

\noindent If $\eta$ is larger than some specified tolerance, the new density
is then mixed with the old density in the form
$n(z)=n_{old}(z)(1-f)+n_{new}(z)f$ where f is a suitably chosen number between
zero and one. This density is
used as input to the calculation and the procedure is iterated until
$\eta$ is smaller than the tolerance. That is, convergence is achieved when the
previous
density and the new density do not vary much.

\indent The above procedure is correct for quantum wells. For a heterostructure
\cite{SDS1}, however, this procedure requires modification since
$m^{*}=m^{*}(z)$ and $\kappa = \kappa(z)$. The Schr\"{o}dinger equation takes
the form

\begin{equation}
\left( -\frac{\hbar^{2}}{2} \frac{d}{dz} \frac{1}{m^{*}(z)} \frac{d}{dz} +
V_{EFF}(z) \right) \xi _{i}(z) = E_{i} \xi _{i}(z)
\end{equation}

\noindent with $V_{EFF}(z)$ still being given by equation\ (\ref{potential})
where $V_{W}$ is

\begin{equation}
V_{W}(z)=V_{0} \theta \left (-z \right) .
\end{equation}

\noindent  Poisson's equation for a position dependent dielectric constant is

\begin{equation}
\frac{d}{dz} \kappa (z) \frac{dV_{H}}{dz}= -4 \pi e^{2} \left( n(z)-n_{I}(z)
\right).
\end{equation}

\noindent The remaining pieces of the self-consistent calculation for
heterostructures are
unchanged from the quantum well case.

\indent The main uncontrolled approximation we are making in applying this LDA
procedure to FQHE calculations is the assumption that the applied external
magnetic field does not appreciably affect the LDA results. Because the applied
magnetic field is in the $z$ direction it is not unreasonable  to assume that
the single particle Schr\"{o}dinger equation in the $z$ variable is not
substantially modified by the magnetic field. But we assume uncritically that
$V_{XC}(z)$ has no
\underline{explicit} magnetic field dependence, which should be a reasonable
approximation for subband quantization arising from $z$ confinement.

\section{Pseudopotentials}

\indent The basic ingredients entering the finite size FQHE diagonalization
study are the Coulomb pseudopotential parameters, $V_{m}$, introduced by
Haldane \cite{HALDANE3,HALDANE4}. Once all the $V_{m}$'s are known, the FQHE
Hamiltonian is completely defined.
The pseudopotential parameters are the energies of pairs
of particles with relative angular momentum $m$. They are given by
\cite{HALDANE4}

\begin{equation}
 V_{m} = \int_{0}^{ \infty } qdq \tilde{V}(q) \left( L_{n}
\left(\frac{q^{2}}{2} \right) \right)^{2} L_{m} \left(q^{2} \right) \exp{
\left(-q^{2} \right)}
\end{equation}

\noindent where $\tilde{V}(q)$ is the Fourier transform of the electron
interaction potential, $V(r)$, and $n$ is the Landau level index.
For small (large) $m$, $V_{m}$ describes the short (long) range part of the
interaction.
If the electrons are fully spin-polarized, then only $V_{m}$ with odd $m$ are
relevant. For the density as determined from an LDA calculation, where
$|\xi (z)|^{2}$ represents the density profile in the $z$ direction, the
relevant equation for $\tilde{V}(q)$ is given by

\begin{equation}
 \tilde{V}(q) = \frac{2 \pi e^{2}}{\kappa q} \int dz_{1} \int dz_{2} | \xi
(z_{1})|^{2} |\xi (z_{2})|^{2}  \exp{(-q|z_{1}-z_{2}|)}.
\end{equation}

\noindent Various approximations to the electron wave function in Eq. (18) give
rise to different pseudopotential parameters $V_{m}$.

\indent The simplest approximation that one can make is to take the
electron-electron interaction to be a pure 2D Coulomb interaction. In this
case,

\begin{equation}
\tilde{V}(q)=\frac{2 \pi e^{2}}{\kappa} \frac{1}{q}.
\end{equation}

\noindent In order to take into account the effect of finite layer thickness
in a quasi-2D electron system, a useful and simple approximation for the
electron-electron interaction is \cite{SDS5} the finite-$\lambda$ model

\begin{equation}
V(\vec{r})=\frac{e^{2}}{\kappa} \frac{1}{ \left(r^{2}+ \lambda ^{2} \right)
^{1/2} }
\end{equation}

\noindent where $\lambda$ is the effective half-width of the electron layer
in the $z$ direction and $\vec{r}$ is the 2D position vector. In momentum
space,

\begin{equation}
\tilde{V}(q)=\frac{2 \pi e^{2}}{\kappa} \frac{ \exp \left(-q \lambda
\right)}{q}.
\end{equation}

\noindent For an infinite barrier square quantum well of width $d$ \cite{SDS3},

\begin{equation}
\tilde{V}(q) = \frac{2 \pi e^{2}}{\kappa q} \frac{1}{(qd)^{2}+ 4 \pi ^{2}}
\left( 3qd +\frac{8 \pi ^{2}}{qd}-\frac{32 \pi ^{4} \left(1-\exp \left(-qd
\right)  \right )}{(qd)^{2}((qd)^{2}+ 4 \pi ^{2})} \right).
\end{equation}

\noindent For a  heterostructure, the Fang-Howard variational result
\cite{ANDO1} is

\begin{eqnarray}
\tilde{V}(q) & = & \frac{2 \pi e^{2}}{\kappa_{avg} q}  \left( \frac{1}{16}
\left(1+ \kappa_{rel} \right) \left( 1 + \frac{q}{b} \right ) ^{-3}
\left(8+\frac{9q}{b}+\frac{3q^{2}}{b^{2}} \right) \right. \nonumber \\
&  & \mbox{} \left. + \frac{1}{2} \left( 1- \kappa_{rel} \right) \left( 1+
\frac{q}{b} \right) ^{-6} \right)
\end{eqnarray}

\noindent with $\kappa_{avg}=(\kappa_{sc} + \kappa_{ins})/2$ being the average
dielectric constant and $\kappa_{rel} = \kappa_{ins} / \kappa_{sc}$ being the
relative dielectric constant of the insulating and semiconductor materials and
$b=3/z_{0}$
where $z_{0}$ is the average extent of the electron wave function in the $z$
direction.
In terms of the density, $b$ is given by

\begin{equation}
b= \left( 48 \pi m^{*} e^{2} N^{*} / \kappa_{sc} \hbar ^{2} \right)^{1/3}
\end{equation}

\noindent where

\begin{equation}
N^{*}= N_{d}+ \frac{11}{32} N_{s}
\end{equation}

\noindent with $N_{d}$ being the depletion charge density in GaAs and $N_{s}$
is the 2D electron density in the layer.

\indent A major focus of our work is to determine how accurate these
approximate models are when used in a FQHE calculation. In particular,
pseudopotential
parameters for these simple model approximations will be compared to those
calculated using the
self-consistent LDA calculation.

\indent The eigenstates of a many-body Hamiltonian are unchanged if the
Hamiltonian (or the potential in the Hamiltonian) is shifted by a constant
amount.
This suggests \cite{SONG} that differences of the $V_{m}$ would be a useful
quantity to
look at. The f-parameters are defined \cite{SONG} in terms of the
pseudopotential
parameters by

\begin{equation}
f_{m} = \frac{V_{3}-V_{m}}{V_{1}-V_{3}}.
\end{equation}

\noindent $f_{1}=-1$ and $f_{3}=0$ for any pair potential.
The Laughlin $\nu=1/3$ state is the exact nondegenerate ground state
for a hard core model Hamiltonian \cite{HALDANE4}. In terms of pseudopotential
parameters, the hard core model
is given by $\{ V_{1},V_{3},V_{5}, \ldots \} = \{V_{1},0,0, \ldots \}$ and
its f-parameters are $\{f_{1},f_{3},f_{5}, \ldots \} = \{-1,0,0, \ldots \}$.
A large deviation from these values implies that the system is not well
represented
by the hard core model and consequently the ground state of the system may not
be incompressible. Our goal in this paper is to investigate the ground state
incompressibility in increasingly more realistic approximations for the Coulomb
pseudopotentials.

\section{Parabolic Quantum Well}

\indent In this section we show the results obtained for a wide parabolic
quantum well. A PQW is constructed by grading the Al concentration
in such a way as to give the conduction band edge a parabolic shape.
As the areal electron density in the well is increased, the half width at
half maximum of the density, $\lambda$, increases. Shayegan {\em et
al.}~\cite{SHAYEGAN},
reported that the FQHE excitation gap decreases dramatically when
$\lambda/l_{c} \approx 3.5 \mbox{ to } 5$. This would indicate that the FQHE
is becoming weakened and that the ground state of the system is no longer
incompressible.

{}From the physical parameters given in Shayegan {\em et al.}~\cite{SHAYEGAN},
we take $V_{0}=276 \mbox{ mev}$, $\alpha= 5.33 \times 10^{-5} \mbox{ mev}/
\mbox{\AA}^{2}$, and $a=3000 \mbox{\AA}$.
The LDA pseudopotential
parameters were calculated using these values for several densities. In
Fig.~\ref{IA}, we show our calculated LDA $V_{m}$ for LDA for the
experimentally determined carrier densities of Shayegan{\em et
al.}~\cite{SHAYEGAN}
compared with the $V_{m}$ for a pure Coulomb interaction.
In Fig.~\ref{IAA}, $V_{m}$ for two relevant approximate models, the infinite
well model
and the finite-$\lambda$ model, are shown. For $m$ greater than approximately
$12$,
$V_{m}$  for the different models agree well with the LDA pseudopotentials.
For small $m$ the pure Coulomb and the infinite well model
seem to overestimate $V_{m}$, while the finite-$\lambda$ model
underestimates $V_{m}$. In Figs.~\ref{IB} and ~\ref{IBB} we show the
corresponding f-parameters
for these pseudopotential parameters. The f-parameters in the finite-$\lambda$
model rise more rapidly with increasing density than the f-parameters
for LDA and the other models. This would give the appearance that for large
densities the ground state would no longer be incompressible in the
finite-$\lambda$ model, as has been concluded ~\cite{SONG} in the literature.

\indent Using the LDA pseudopotential parameters, we studied the $\nu=1/3$ FQHE
state employing the finite size exact diagonalization technique.
Figs.~\ref{IC} and~\ref{ID} show the calculated overlap with the Laughlin
$\nu=1/3$
state, and the calculated bare excitation gap, $\Delta$, and the gap
minus the level broadening, $\Delta -2 \Gamma$, as a function of electron
density.
Also shown is the gap as measured experimentally \cite{SHAYEGAN} by Shayegan
{\em et al.} The agreement
between the experimental results and our calculation is very good. For a pure
Coulomb interaction,
$\Delta \approx 14.2 \mbox{ K}$ and $\Delta -2 \Gamma \approx 11.4 \mbox{ K}$
and using
the finite-$\lambda$ model, $\Delta \approx 2.9 \mbox{ K to } 1.4 \mbox{ K}$
and $\Delta -2 \Gamma \approx 0.1 \mbox{ K to } 0.0 \mbox{ K}$.
The overlap of the LDA result with the Laughlin state is found to be quite
large for all
densities. This is to be contrasted with the finite-$\lambda$ model where
the overlap is $ \approx 0.8 \mbox{ to } 0.4$ for the given range of densities.
For a pure Coulomb interaction the overlap with the Laughlin state is also
quite large.

\indent We also studied the subband dependence of the PQW results in an
artificial model calculation which bears no resemblance to reality.
Pseudopotential parameters were calculated assuming that (a) only the lowest
subband was occupied, (b) only the first excited subband was occupied, and (c)
only
the second excited subband was occupied. We then performed a FQHE calculation
using
the parameters for (a)-(c). The difference between these results and our full
LDA results was quite small (less than $1 \%$) for the overlap with the
Laughlin
$\nu=1/3$ state. The differences for the gap were larger (as much as $30 \%$).

\section{Heterostructure}

\indent For the LDA calculation, we took the physical parameters to be those
appropriate for a typical GaAs-$\mbox{Al}_{x} \mbox{Ga}_{1-x}$As
heterostructure:
$V_{0}=276 \mbox{ mev}$, $\kappa_{sc} = 12.8$, $\kappa_{ins}=12.1$,
$m^{*}_{sc}=0.068 \mbox{ } m_{0}$, and $m^{*}_{ins}=0.088 \mbox{ } m_{0}$. In
Fig.~\ref{IIA}  we
show the pseudopotential parameters for electron densities between $1 \times
10^{10}$
 to $3 \times 10^{11}  \mbox{cm}^{-2}$. From Fig.~\ref{IIAA} it is clear that
the approximate variational model
 \cite{ANDO1} is fairly reliable when compared to the LDA
pseudopotentials especially for large densities.
Figs.~\ref{IIB} and~\ref{IIBB} show the corresponding f-parameters. These
parameters are
fairly constant for all densities greater than $ \approx 1 \times 10^{11}
\mbox{cm}^{-2}$.
For smaller densities, the f-parameters deviate more strongly from the hard
core model f-parameters.

The overlap with the Laughlin state as a function of density is
shown in Fig.~\ref{IIC}. As expected it is quite large, especially for the
larger
densities. The finite-$\lambda$ model also gives a large overlap
with the Laughlin state, $ \approx 0.9 \mbox{ to } 0.99$ for this range of
densities. The disagreement between the LDA results and the finite-$\lambda$
model is larger for smaller densities.

Figure~\ref{IID} shows calculated gaps $\Delta$ and
$\Delta - 2 \Gamma$ as a function of density. The agreement between
our `subtracted' gap and the experimental measurement \cite{WILLET1} of Willet
{\em et al.} is
very good. For comparison, a pure Coulomb interaction gives $\Delta \approx
14.2 \mbox{ K}$ and $\Delta -2 \Gamma \approx 13.8 \mbox{ K}$, while the
finite-$\lambda$ model gives $\Delta \approx 12.8 \mbox{ K to } 5.7 \mbox{ K}$
and $\Delta -2 \Gamma \approx 12.7 \mbox{ K to } 5.3 \mbox{ K}$.

\section{Square Quantum Well}

\indent  We consider a typical narrow SQW with a width $139 \mbox{\AA}$ and
$V_{0}=276 \mbox{ mev}$,
for a typical range of densities, $N_{S} = 1 \times 10^{10} \mbox{ to } 5
\times 10^{11} \mbox{cm}^{-2}$.
The LDA pseudopotentials are shown in Fig.~\ref{IIIA}. For this well and this
range
of densities the calculated $V_{m}$ show very little density variation. The
pseudopotential parameters for the other models are shown in Fig.~\ref{IIIAA}.
As in
the PQW case, the infinite well model overestimates the $V_{m}$ while
the finite-$\lambda$ model underestimates the $V_{m}$. However, for all
$m$ greater than 4, the $V_{m}$ for all of the models are approximately
equal. The f-parameters are shown in Figs.~\ref{IIIB} and~\ref{IIIBB}. These
parameters for
the LDA model are almost constant for the given range of densities and they
are close to being equal in all of the models. These parameters do not
rise above one for any of the models. Since the $V_{m}$s remain small for all
$m$, it is reasonable to assume that the ground state
should be incompressible over this range of densities in a square quantum well.

\indent The overlap of the exact numerical wave function with the $\nu=1/3$
Laughlin state is shown in Fig.~\ref{IIIC}. It also shows almost no variation
with
density and it is very close to the overlap computed using either the
pure Coulomb or the finite-$\lambda$ model.

\indent The gap, as shown in Fig.~\ref{IIID}, shows almost no variation with
density.
It is very close to the pure Coulomb value, $\Delta \approx 14.2 \mbox{ K}$.
Using the finite-$\lambda$ model, $\Delta \approx 13.8 \mbox{ K}$ for this
range of densities. Thus, for a square quantum well, all of the approximations
for the pseudopotential should work well in FQHE calculations.

\section{Even denominator FQHE : $\nu=5/2$}

\indent The first unambiguous observation of an even denominator filling factor
in a single layer system was made by Willet {\em et al.} \cite{WILLET2}.
Magnetotransport experiments
carried out in a high mobility GaAs-AlGaAs heterostructure showed
\cite{WILLET2} a plateau in the
Hall resistivity concurrent with a deep minima in the longitudinal
resistivity corresponding to a filling factor of $5/2$. Tilted field
experiments \cite{EISEN1} on the $5/2$ state have shown that it is rapidly
destroyed by increasing the Zeeman energy
which indicates that the spin degree of freedom may be important in
understanding this state.

\indent In analogy with the Laughlin state for odd denominator filling
factors, Haldane and Rezayi \cite{HALDANE6} have proposed a `hollow core' model
wave function
that may describe the physics of the $\nu=5/2$ state. This
spin-singlet wave function represents an incompressible state for $\nu=1/2$.
However this hollow-core wave function requires a substantially reduced short
range repulsion between
the electrons relative to a pure 2D Coulomb interaction.

\indent From the physical parameters for the heterostructure of Willet {\em
et al.} \cite{WILLET2} we calculate the LDA pseudopotential parameters for the
first
Landau level ($n=1$). These parameters are shown in Table 1 and in
Fig.~\ref{IVA}.
Figure~\ref{IVA} also shows the effect on the pseudopotential parameters for
the
variational model when the
electron density is varied by 20\% and also the effect of varying the relative
dielectric constant, $\kappa_{ins}/\kappa_{sc}$ from $0$ to $1.5$. Our reason
for varying the system parameters ({\em i.e.} electron density and background
dielectric constants) is to check whether such parameter modifications could
produce an incompressible hollow core state at $\nu=5/2$. Assuming
that the lowest Landau level is completely filled and inert, we perform
a finite size diagonalization calculation for a system of eight electrons
in the spherical geometry. Shown in Fig.~\ref{IVB} is the excitation spectrum
as a
function of total angular momentum $L$ for total spin $S$ equal to 4. We find
that the ground state is in the $L=0$ $S=4$ sector. The ground state energy
for $L=0$ $S=4$ is however close to the energies found for the other $L=0$
sectors.
The overlap of our
wave function from the finite size diagonalization calculation with the hollow
core model is quite small ($5 \times 10^{-3}$) indicating that this model is
not a good candidate for the $5/2$ state. At this stage, therefore, we
conclude, in agreement with earlier investigations \cite{MAC1} of this issue,
that the $5/2$ FQHE phenomenon as observed in ref.~\onlinecite{WILLET2} remains
unexplained
theoretically, and in particular, the hollow core model proposed in
ref.~\onlinecite{HALDANE6} is not quantitatively consistent with the system
parameters of the experimental
sample in ref.~\onlinecite{WILLET2}.

\section{Conclusion}
\indent In summary, we have obtained realistic Coulomb pseudopotential
parameters for FQHE calculations in 2D GaAs--$\mbox{Al}_{x} \mbox{Ga}_{1-x}$As
quantum structures using a self-consistent LDA electronic subband structure
results. We compare the LDA pseudopotential parameters with those from a number
of simpler model approximations ({\em eg.} the pure 2D Coulomb model, the
finite-$\lambda$
 model, the infinite well model, and the variational model) to
estimate the quantitative accuracy of the simpler models for various systems
and different
electron densities. Our most realistic calculations yield FQH excitation gaps
which, when corrected for the level broadening effect, are in excellent
quantitative agreement with the experimentally determined activation gaps as
obtained from transport data. For the $\nu=5/2$ FQHE observed in
ref.~\onlinecite{WILLET2} our
calculations show that the hollow core model of ref.~\onlinecite{HALDANE6} is
quantitatively
inconsistent with the LDA Coulomb pseudopotentials for the experimental
sample parameters of ref.~\onlinecite{WILLET2}. Our calculated realistic
Coulomb pseudopotentials
for various systems should enable future FQHE finite size exact diagonalization
 calculations to be quantitatively more realistic.

\indent This work is supported by the US-ONR.

\begin{figure}[p]
\caption{Pseudopotential parameters in units of $e^{2}/\kappa l_{c}$ for a pure
Coulomb interaction (short lines) and for the LDA calculation (long lines) for
a PQW. The electron densities are: $4.9, 6.0, 7.3, 8.5 \times 10 ^{10}
\mbox{cm}^{-2}$.} \label{IA}
\end{figure}

\begin{figure}[p]
\caption{Pseudopotential parameters in units of $e^{2}/\kappa l_{c}$ for the
finite-$\lambda$ model (short lines) and the infinite well model (long lines)
for a PQW. The electron densities are the same as in Fig. 1.} \label{IAA}
\end{figure}

\begin{figure}[p]
\caption{f-parameters as a function of electron density for the LDA
pseudopotential parameters for a PQW for odd $m$.} \label{IB}
\end{figure}

\begin{figure}[p]
\caption{f-parameters as a function of electron density for the
finite-$\lambda$ model (dashed lines) and the infinite well model (solid lines)
for a PQW for odd $m$.} \label{IBB}
\end{figure}

\begin{figure}[p]
\caption{Overlap between the Laughlin $\nu=1/3$ state and the exact numerical
ground state found using the LDA pseudopotential parameters for a PQW.}
\label{IC}
\end{figure}

\begin{figure}[p]
\caption{The excitation gap (dashed line), the `subtracted gap' (solid line),
and the experimental measurement~\protect\cite{SHAYEGAN} of Shayegan {\em et
al.} ($ \ast $) for a PQW.} \label{ID}
\end{figure}

\begin{figure}[p]
\caption{Pseudopotential parameters in units of $e^{2}/\kappa l_{c}$ for the
pure Coulomb (short lines) and the LDA results (long lines) for a
heterostructure. The electron densities are $0.1, 0.5, 1.0, 2.0, 3.0 \times
10^{11} \mbox{cm}^{-2}$.} \label{IIA}
\end{figure}

\begin{figure}[p]
\caption{Pseudopotential parameters in units of $e^{2}/\kappa l_{c}$ for a
heterostructure for the variational model (long lines) and the finite-$\lambda$
model (short lines). The electron densities are the same as in Fig. 7.}
\label{IIAA}
\end{figure}

\begin{figure}[p]
\caption{f-parameters as a function of electron density for the LDA
pseudopotential parameters for a heterostructure for odd $m$.} \label{IIB}
\end{figure}

\begin{figure}[p]
\caption{f-parameters as a function of electron density for the the variational
model (solid lines) and the finite-$\lambda$ model (dashed lines) for a
heterostructure for odd $m$.} \label{IIBB}
\end{figure}

\begin{figure}[p]
\caption{Overlap between the Laughlin $\nu=1/3$ state and the exact numerical
ground state found using the LDA pseudopotential parameters for a
heterostructure.} \label{IIC}
\end{figure}

\begin{figure}[p]
\caption{The excitation gap (dashed line), the `subtracted' gap (solid line),
and the experimental measurement~\protect\cite{WILLET1} of Willet {\em et al.}
($ \ast $) for a heterostructure.} \label{IID}
\end{figure}

\begin{figure}[p]
\caption{Pseudopotential parameters in units of $e^{2}/\kappa l_{c}$ for the
pure Coulomb interaction (short lines) and the LDA results (long lines) for a
SQW. The electron densities are $0.1, 0.5, 1.0, 5.0 \times 10^{11}
\mbox{cm}^{-2}$.} \label{IIIA}
\end{figure}

\begin{figure}[p]
\caption{Pseudopotential parameters in units of $e^{2}/\kappa l_{c}$ for the
finite-$\lambda$ model (short lines) and the infinite well model (long lines)
for a SQW. The electron densities are the same as in Fig. 13.} \label{IIIAA}
\end{figure}

\begin{figure}[p]
\caption{f-parameters as a function of electron density for the LDA
pseudopotential parameters for a SQW for odd $m$.} \label{IIIB}
\end{figure}

\begin{figure}[p]
\caption{f-parameters as a function of electron density for the
finite-$\lambda$ model (dashed lines) and the infinite well model (solid lines)
for a SQW for odd $m$.} \label{IIIBB}
\end{figure}

\begin{figure}[p]
\caption{Overlap between the Laughlin $\nu=1/3$ state and the exact numerical
ground state found using the LDA pseudopotential parameters for a SQW.}
\label{IIIC}
\end{figure}

\begin{figure}[p]
\caption{The excitation gap as a function of electron density for a SQW.}
\label{IIID}
\end{figure}

\begin{figure}[p]
\caption{LDA pseudopotential parameters in units of $e^{2}/\kappa l_{c}$ for
the heterostructure of Willet {\em et al.}~\protect\cite{WILLET2} (long lines)
for the $\nu=5/2$ FQHE. Also shown are the pseudopotential parameters for the
variational model (short lines) assuming a twenty percent variation in the
electron density and separately assuming $\kappa_{rel}$ is between $0$ and
$1.5$.} \label{IVA}
\end{figure}

\begin{figure}[p]
\caption{The excitation spectrum for $S=4$ for an eight particle calculation
using the LDA pseudopotential parameters for the heterostructure of Willet {\em
et al.}~\protect\cite{WILLET2}}
\label{IVB}
\end{figure}

\begin{table}[p]
\caption{The LDA pseudopotential parameters in units of $e^{2}/\kappa l_{c}$
for the heterostructure of Willet {\em et al.} ~\protect\cite{WILLET2}}
\label{IVDD}
\end{table}

\vspace{5cm}

\begin{tabular}{||l||l||r||}      \hline
   m   &  $V_{\mbox{m}}$    \\ \hline
   0   &  0.47665957508     \\ \hline
   1   &  0.37332084804     \\ \hline
   2   &  0.35230370587     \\ \hline
   3   &  0.28844405415     \\ \hline
   4   &  0.24996235081     \\ \hline
   5   &  0.22361694066     \\ \hline
   6   &  0.20414646380     \\ \hline
   7   &  0.18900642725     \\ \hline
   8   &  0.17679961714     \\ \hline
   9   &  0.16668746764     \\ \hline
  10   &  0.15813248355     \\ \hline
  11   &  0.15077224603     \\ \hline
  12   &  0.14435235363     \\ \hline
  13   &  0.13868828532     \\ \hline
  14   &  0.13364252111     \\ \hline
  15   &  0.12911019556     \\ \hline
  16   &  0.12500975822     \\ \hline
  17   &  0.12127670242     \\ \hline
  18   &  0.11785925293     \\ \hline
  19   &  0.11471532800     \\ \hline
  20   &  0.11181031734     \\ \hline
\end{tabular}

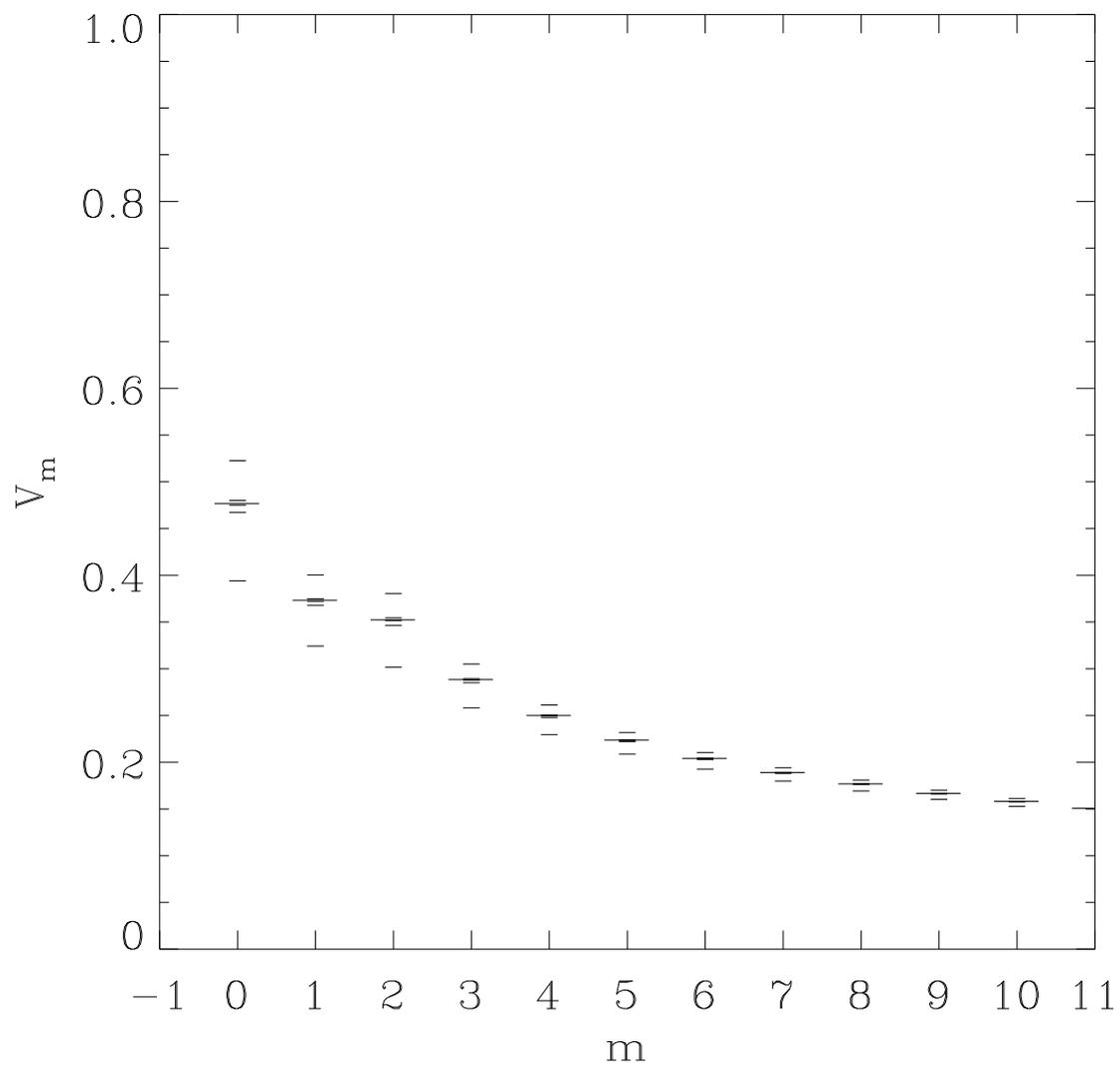
\end{document}